# MAXIMAL FREQUENT ITEMSET GENERATION USING SEGMENTATION APPROACH

M.Rajalakshmi[1], Dr.T.Purusothaman[2], Dr.R.Nedunchezhian[3]

[1]Assistant Professor (SG), Coimbatore Institute of Technology, India,
`rajalakshmi@cit.edu.in`
[2]Associate Professor, Government College of Technology, India,
`purushgct@yahoo.com`
[3]Vice-Principal, Kalaignar Karunanidhi Institute of Technology, India
`rajuchezhian@gmail.com`

## ABSTRACT

*Finding frequent itemsets in a data source is a fundamental operation behind Association Rule Mining. Generally, many algorithms use either the bottom-up or top-down approaches for finding these frequent itemsets. When the length of frequent itemsets to be found is large, the traditional algorithms find all the frequent itemsets from 1-length to n-length, which is a difficult process. This problem can be solved by mining only the Maximal Frequent Itemsets (MFS). Maximal Frequent Itemsets are frequent itemsets which have no proper frequent superset. Thus, the generation of only maximal frequent itemsets reduces the number of itemsets and also time needed for the generation of all frequent itemsets as each maximal itemset of length m implies the presence of $2^m$-2 frequent itemsets. Furthermore, mining only maximal frequent itemset is sufficient in many data mining applications like minimal key discovery and theory extraction. In this paper, we suggest a novel method for finding the maximal frequent itemset from huge data sources using the concept of segmentation of data source and prioritization of segments. Empirical evaluation shows that this method outperforms various other known methods.*

## KEYWORDS

*Knowledge discovery in data sources, maximal frequent itemset, association rules, data mining, segmentation*

## 1. INTRODUCTION

Data mining plays an important role to retrieve valuable, hidden and predictive information from huge data sources. It is a powerful technology with great potential to analyze important information which can be used for strategic decision making. It is an essential step in Knowledge Discovery in Databases (KDD) which is the process of identifying valid, potentially useful and ultimately understandable patterns in data. (Han and Kamber, 2001). The various steps in knowledge discovery process include data cleaning, data integration, data selection, data transformation, data mining, pattern evaluation and knowledge presentation.

Data mining functionalities like association rule mining, cluster analysis, classification and prediction etc. are used to specify the kind of patterns to be mined. Among the functionalities of data mining, Association Rule Mining (ARM) has received a great deal of attention and it finds interesting associations or correlation relationships among a large set of data items. Finding association rules among huge amount of business transaction records can help in many business decision making processes such as catalog design, cross marketing, etc. The best example of ARM is market basket analysis which involves analyzing the customer buying habits from the association between the different items which is available in the shopping baskets. This analysis can help retailers to develop marketing strategies.

ARM involves two stages:

i. Finding frequent itemsets – By definition, each of these itemsets will occur at least as frequently as a pre-determined minimum support count.
ii. Generating strong association rules – By definition, these rules must satisfy the minimum support and minimum confidence.

Many algorithms have been proposed for frequent itemsets generation. They are Apriori(Agrawal and Srikant, 1994), Pincer search (Lin and Kedem, 1998 ,2002), Frequent Pattern tree (Han et al., 2004), etc. Some applications do not require all the frequent itemsets of varying lengths as they are voluminous and laborious to generate. Hence, a reduced set called the Maximal Frequent itemset is generated. Maximal Frequent Itemsets are frequent itemsets which have no proper frequent superset.

The rest of the paper is organized as follows. Section 2 gives the problem definition. Section 3 describes the related existing work. The proposed approach is explained in Section 4. Section 5 compares the performance of the proposed algorithm with Pincer technique and Section 6 concludes the paper.

## 2. PROBLEM DEFINITION

Let $I = \{ i_1, i_2, \ldots i_m\}$ be the set of items and D be the transactional data source which contains the set of transactions. Each transaction T is a set of items such that $T \subseteq I$ and is associated with an identifier called TID. An association rule is an implication of the form X=>Y, where $X \subset I$, $Y \subset I$ and $X \cap Y = \Phi$. In general, every association rule must satisfy two user specified constraints, one is support($\sigma$) and the other is confidence ($\Gamma$). The support of a rule X=>Y is defined as the fraction of transactions that contain X∩Y, while the confidence is defined as the ratio of support(X∩Y)/support(X). An itemset is frequent if its support satisfies at least the minimum support, otherwise it is said to be infrequent. A frequent itemset is a Maximal Frequent itemset if it is a frequent set and no superset of this is a frequent set. The paper aims to find the Maximal Frequent itemset from a huge data source.

## 3. RELATED WORK

The problem of generating frequent itemsets is fundamental for many data mining tasks, such as mining association rules (Agrawal and Srikant, 1994), correlations, sequential patterns (Agrawal and Srikant, 1995), associative classification (Liu et al., 1998) etc. Many algorithms have been proposed to find Maximal Frequent itemset namely MaxEclat, MaxClique (Zaki, 2000), Pincer search (Lin and Kedem, 1998 ,2002), Maxminer (Bayardo, 1998), Depth project (Agrawal et al., 2000), Mafia (Burdick et al., 2001), GenMax (Gouda and Zaki, 2005) and FPMax (Grahne and Zhu, 2005).

When the algorithms MaxEclat, MaxClique, Pincer search, Maxminer were proposed, size of the main memory was small and the whole database could not be held in main memory, therefore multiple scans of database was needed. Several algorithms were designed to reduce the number of scans. However, there are some small size dense data sources from which large number of maximum frequent itemsets are generated. For this purpose more main memory space is needed during processing.

In addition, when the length of frequent pattern is long, the number of maximum frequent itemset increases exponentially which makes the problem computationally difficult. Even though the main memory size has been increased in the past few decades, it is reasonable to assume that some databases can be held into main memory and to therefore focus on CPU efficiency in algorithm design. Algorithms DepthProject, MAFIA, Genmax and FPmax make this assumption.

Bayardo (1998) introduces MaxMiner which extends the Apriori algorithm to mine only the maximal frequent itemsets. To reduce the search space, MaxMiner performs not only subset infrequency pruning, but also a "lookahead" to do superset frequency pruning. Though superset frequency pruning reduces the search time dramatically, MaxMiner still needs many passes to get all maximal frequent itemsets. Zaki et al. [max Eclat and max clique] present the algorithms MaxEclat and Maxclique for identifying MFS. These algorithms are similar to MaxMiner in that they also attempt to look ahead and identify MFS early. The important difference is that MaxMiner attempts to look ahead throughout the search, whereas MaxEclat and Maxclique look ahead only during an initialization phase prior to a purely bottom-up Apriori-like search with exponential scaling. The initialization phase of MaxClique is also prone to problems with MFS since it uses a dynamic programming algorithm to find maximal cliques in a graph whose largest clique is at least as large as the length of the Maximal frequent itemsets.

DepthProject by Agarwal et al., (2000) also mines only maximal frequent itemsets. It performs a mixed depth-first and breadth-first traversal of the itemset lattice. In the algorithm, both subset infrequency pruning and superset frequency pruning are used. The database is represented as a bitmap. Each row in the bitmap is a bitvector corresponding to a transaction and each column corresponds to an item. The number of rows is equal to the number of transactions, and the number of columns is equal to the number of items. By using the carefully designed counting methods, the algorithm significantly reduces the cost for finding the support counts.

The paper (Burdick et al., 2001) extends the idea in DepthProject and give an algorithm called Mafia to mine the maximal frequent itemsets. Similar to DepthProject, their method also uses a bitmap representation, where the count of an itemset is based on the column in the bitmap. Mafia is a depth-first algorithm which has besides subset infrequency pruning and superset frequency pruning, another pruning technique called Parent Equivalence Pruning Technique.

Lin and Kedem (1998, 2002) proposed the Pincer algorithm which combines both bottom-up and top-down searches to identify frequent itemsets effectively. All the itemsets are not explicitly examined. It classifies the data source into three classes as frequent, infrequent, and unclassified data. Bottom-up approach is the same as Apriori. Top-down search uses a new set called Maximum-Frequent-Candidate-Set (MFCS). It also uses another set called the Maximum Frequent Set (MFS) which contains all the maximal frequent itemsets identified during the process. Any itemset that is classified as infrequent in bottom-up approach is used to update MFCS. Any itemset that is classified as frequent in the top-down approach is used to reduce the number of candidates in the bottom–up approach. When the process terminates, both MFCS and MFS are equal. This algorithm involves more data source scans in the case of sparse data sources.

Since the performance of the proposed approach is compared with pincer search algorithm, the working principle of pincer search is elaborated in the following section.

### 3.1 PINCER SEARCH ALGORITHM

Most of the algorithms used for mining maximal frequent itemsets perform fairly well when the length of the maximal frequent itemset is small. However, performance degrades when the length of the maximal frequent itemset is large, since in the bottom-up approach, the maximal frequent itemset is obtained only after traversing all its subsets.

The Pincer-search algorithm (Lin and Kedem, 1998, 2002), proposes a new approach for mining maximal frequent itemsets. It reduces the complexity by combining both top-down and bottom-up methods for generating maximal itemsets. The bottom-up search starts from 1-itemset and proceeds upto *n*-itemsets as in Apriori while the top-down search starts from n-itemsets and proceeds upto 1-itemset. Both bottom-up and top-down searches identify the maximal frequent itemsets by examining its candidates individually. Bottom-up search moves one-level up during a single pass whereas top-down search moves many levels down during a single pass.

During the execution, all the itemsets are classified into 3 categories:

*Frequent:* Itemsets whose support is greater than min_sup are classified as frequent.
*Infrequent*: Itemsets whose support is less than min_sup are classified as infrequent
*Unclassified:* All other itemsets are said to be unclassified

Pincer algorithm uses the following two properties to classify the unclassified itemsets.

> *Property 1:* If an itemset is infrequent, all its supersets must be infrequent and they
>     need not be examined further.
> *Property 2:* If an itemset is frequent, all its subsets must be frequent and they  need
>     not be examined further.

Property 1 is used in the bottom-up approach and Property 2 is used in the top-down search. Pincer search uses both the properties in order to reduce the number of candidates as well as the number of scans over the data source. In this algorithm, top-down search uses a new set called Maximum-Frequent-Candidate-Set (MFCS) which is defined as follows:

The MFCS is a minimum cardinality set of itemsets so that the union of all the subsets of its elements contains all the frequent itemsets but does not contain any infrequent itemsets, i.e., it is a minimum cardinality set satisfying the conditions.

$$\text{FREQUENT} \subseteq \cup \{2^x | X \in MFCS\}$$
$$\text{INFREQUENT} \cap \{2^x | X \in MFCS\} = \Phi$$

where FREQUENT and INFREQUENT, stand for all frequent and infrequent itemsets classified so far respectively.

The Pincer search also uses another set called Maximal Frequent Set (MFS) which contains all maximal frequent itemsets identified so far. During the execution of the process, MFCS will be the superset of MFS. At the termination of the process, both MFCS and MFS are equal.

Initially MFCS contains all the items in the data source; say $(i_1, i_2, ..., i_n)$ where $n$ is the number of distinct items in data source. Bottom-up approach starts from 1-itemsets. Any itemset that is classified as infrequent in bottom-up approach is used to update the MFCS. MFCS is updated by reducing those itemsets in MFCS containing that infrequent itemset as its subset and a new MFCS is formed. All the frequent itemsets in MFCS are added to MFS. Similarly, any itemset that is classified as frequent in the top-down approach is used to reduce the number of candidates in the bottom-up direction. Thus, the information obtained in one direction is used for reducing the number of candidates in the opposite direction.

During the formation of $L_k$, those itemsets in $C_k$ which are subsets of itemsets in MFS are pruned. However, there may be some itemsets which may be subsets of MFS but are required to generate candidates for $C_{k+1}$. Those itemsets are recovered to generate candidates for $C_{k+1}$. This can be done by using a recovery procedure.

The Pincer algorithm is illustrated for the sample data source given in Table 1. The same data source will be used later for illustrating the proposed algorithm. An example of pincer search is shown in figure 1.

**Table 1  Sample data source**

| TID | ITEMS |
|---|---|
| 100 | a d e g j |
| 200 | a b |

| 300 | a b c e h |
| 400 | a b c d |
| 500 | a b c d f i |
| 600 | a b c |
| 700 | a b c d f i |
| 800 | a b c e |
| 900 | j |

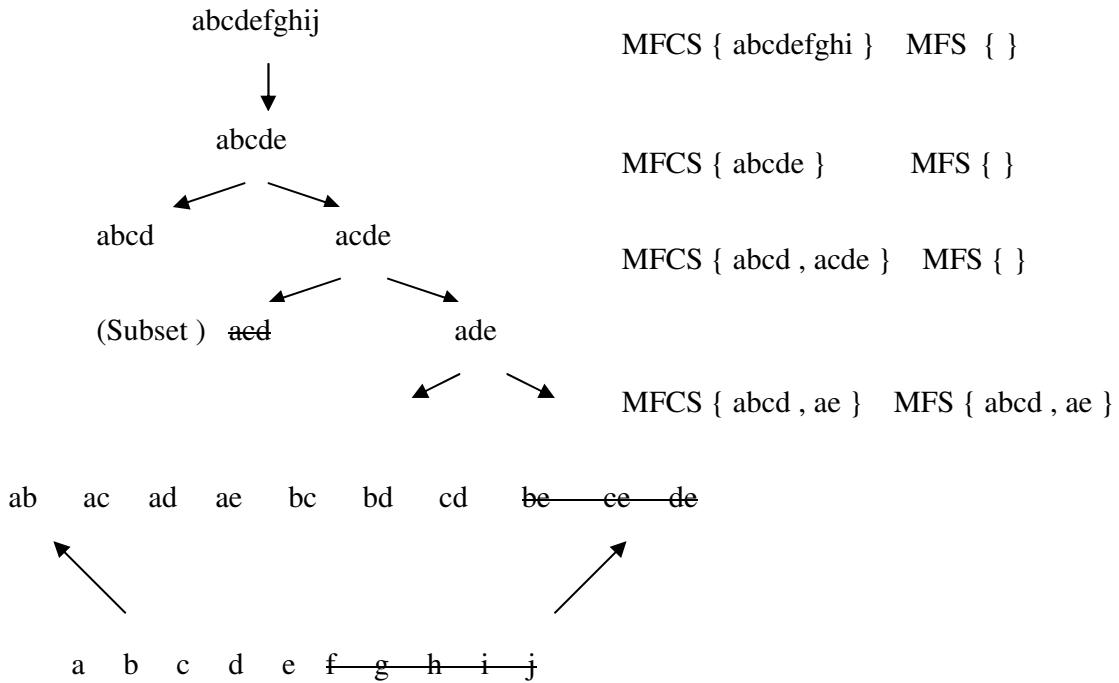

**Figure 1  Pincer search**

Thus, this algorithm reduces the number of candidates and the number of scans, which in turn reduce I/O time and CPU time.

However, in case of some sparse data sources, the performance is degraded as there may not be any information to share between two ends resulting in a larger number of candidates. This causes repeated scans over the data source and the performance degrades to that obtainable by the Apriori algorithm. In this paper, a novel approach is introduced to generate maximal frequent itemset from a sparse data source.

## 4. THE PROPOSED TECHNIQUE

The proposed work describes a novel method for generation of the maximal frequent itemsets with minimum effort. Instead of generating candidates for determining maximal frequent itemsets as done in other methods (Lin and Kedem, 2002), this method uses the concept of partitioning the data source into segments and then mining the segments for maximal frequent itemsets. It thus reduces the number of scans over the transactional data source to merely two. Moreover, the time spent for candidate generation is eliminated.

This algorithm involves the following steps to determine the MFS from a data source.
1. Segmentation of the transactional data source.
2. Prioritization of the segments
3. Mining of segments

The overview of the proposed work is illustrated in Figure 2

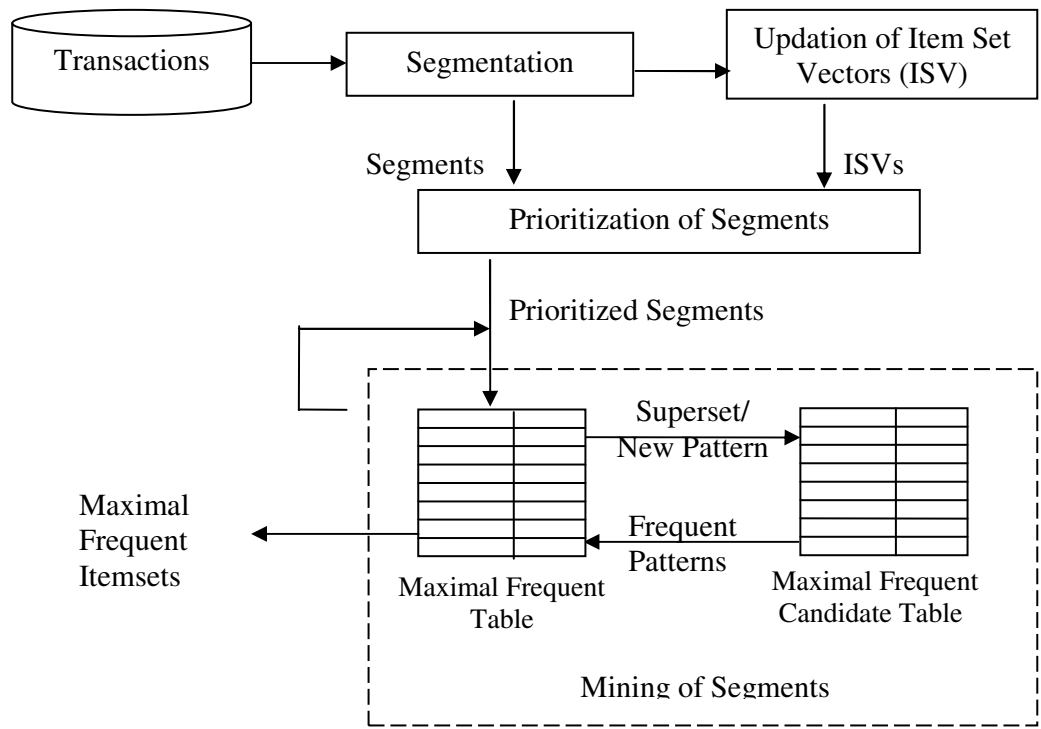

**Figure 2 Overview of the proposed technique**

International Journal of Database Management Systems ( IJDMS ), Vol.3, No.3, August 2011

Segmentation involves dividing the data source into a number of equal-sized segments. After segmenting, the segments are prioritized based on its support count (horizontal count). Once the priority is set, the segments are mined for maximal itemsets in the order of their priority. These steps are discussed in detail below.

### 4.1 SEGMENTATION OF THE TRANSACTIONAL DATA SOURCE

The transactional data source is divided into a number of equal-sized segments. The segment size can be determined depending on the minimum threshold, *min_sup*.

$$\text{Number of equal-sized segments} \leq \frac{100}{\text{min\_sup }\%} \quad (4.1)$$

With such partitioning, the number of transactions in a segment will not be less than min_sup % of the total transactions, |D|. Each segment is associated with a new data structure called Item Support Vector (ISV) kept in main memory which contains the support of each individual item in that segment. The structure of an ISV is given in Table 2.

**Table 2 Item support vector (ISV)**

| Item  | $I_1$ | $I_2$ | … | $I_n$ |
|-------|-------|-------|---|-------|
| Count |       |       |   |       |

For example by segmenting the data source given in Table 1 with each segment size=3, the ISV for segment1 is given in Table 3.

**Table 3 Item support vector for segment 1**

| Item  | a | b | c | d | e | F | g | h | i | j |
|-------|---|---|---|---|---|---|---|---|---|---|
| Count | 3 | 2 | 1 | 1 | 2 | 0 | 1 | 1 | 0 | 1 |

During the first scan of the data source, all the ISVs are filled up. The counts of individual items in a segment are recorded in the appropriate segment's ISV. Finally, the overall support of each item can be calculated from the contents of all ISVs. For example, the overall support of an item '*a*' can be calculated by adding the counts of the item '*a*' in all ISVs. For this example, the support of '*a*' is 8. Only those items having its overall count greater than the specified minimum support are considered for mining MFS.

### 4.2 PRIORITIZATION OF SEGMENTS

Unlike other approaches, the data source is not scanned sequentially for identifying the MFS. Here the proposed method makes use of the contents of ISVs to guide the search selectively. The contents of ISVs reveal the possible combination of items in their respective segments. Before initiating the second scan, the sum of count values of all the frequent items in each segment is calculated. It is called the horizontal count, $h_i$, where i=1,2,...,n. Horizontal count

26

International Journal of Database Management Systems ( IJDMS ), Vol.3, No.3, August 2011

for a segment is calculated by adding the counts of different items in its ISV excluding the infrequent items. Then, these horizontal count values are arranged in descending order and the segments are then prioritized. The segment having the highest horizontal count value is given the highest priority. A segment with highest priority is considered first for mining. A segment with second highest priority is considered next and the mining continues until all the maximal frequent itemsets are identified.

As an illustrative example, the same data source given in the Table 1 is taken. It is divided into three segments. The transactions of the data source are from an ordered domain I= {a, b, c, d, e, f, g, h, i, j}. The minimum support is kept as 3. Therefore only the items a, b, c, d, and e are used to calculate $h_i$. The ISVs and the horizontal and vertical counts for the data source given in Table 1 with its segment size equal to 3 are given in Table 4.

**Table 4 Item support vectors of the data source**

| | Item | | | | | | | | | | |
|---|---|---|---|---|---|---|---|---|---|---|---|
| Segment | a | b | c | d | E | f | g | h | i | j | Horizontal count $h_i$ |
| ISV1 | 3 | 2 | 1 | 1 | 2 | 0 | 1 | 1 | 0 | 1 | 9 |
| ISV2 | 3 | 3 | 3 | 2 | 0 | 1 | 0 | 0 | 1 | 0 | 11 |
| ISV3 | 2 | 2 | 2 | 1 | 1 | 1 | 0 | 0 | 1 | 1 | 8 |
| Vertical count(support) | 8 | 7 | 6 | 4 | 3 | 2 | 1 | 1 | 2 | 2 | |

The data source is scanned in the order of Segment2, Segment1, and Segment3 since their horizontal counts of ISVs are 11, 9, and 8 respectively. The purpose of segment based scanning is to concentrate on the dense areas of the data source to look for MFS.

### 4.3 MINING OF SEGMENTS

The mining process uses two table data structures to locate MFS called *Maximal Frequent Table* and *Maximal Frequent Candidate Table* respectively. Each table maintains patterns and their corresponding counts. The transactions are read from the data source one by one. The frequent items in a transaction called *pattern* alone are considered and recorded in the tables. Mining can be done in the data source as follows:

**I** Start scanning the segments in the order of their priority i.e., descending order of $h_i$.
For each segment apply the following

1. Read a transaction *t*, when the segment is not empty

2. Consider frequent items alone (pattern)

3. Check for the pattern in *Maximal Frequent Table*
    a. If *Maximal Frequent Table* is null, goto step 4.
    b. If the new pattern is a subset of an already existing pattern then goto step 1.
    c. If the new pattern is similar to an already existing pattern, then increment the count of the already existing pattern by one and goto step 1.
    d. If the new pattern is a superset of an already existing pattern then increment the count of the existing pattern by one and goto step 4.

4. Check for the pattern in *Maximal Frequent Candidate Table*
    a. If the incoming pattern is similar to an existing pattern then increment its count by one
    b. If the incoming pattern is superset of an existing pattern then increment the count of its sub pattern by one and insert the new pattern into *Maximal Frequent Candidate Table* with its count initialized to 1.
    c. If the incoming pattern is a subset of an existing pattern then insert the new pattern into *Maximal Frequent Candidate Table* with its count initialized to the sum of the supports of its supersets in *Maximal Frequent Candidate Table* plus one.
    d. If the incoming pattern is a new pattern, record the new pattern in the *Maximal Frequent Candidate Table* and initialize its count to one.

5. If any of the patterns in the *Maximal Frequent Candidate Table* has reached the minimum support threshold, move that pattern to *Maximal Frequent Table*. Simultaneously prune all its sub patterns from the *Maximal Frequent Table*.

**II** Determine all the common frequent patterns from the patterns in *Maximal Frequent Candidate Table* and calculate its support which is equal to the sum of the support of its supersets in *Maximal Frequent Candidate Table*. Add it to the *Maximal Frequent Table* when its support passes the minimum threshold.

**Example**

For the data source given in Table 1 mining is done as follows.

Step I:

After mining segment 2,
    *Maximal Frequent Table* contains {*abc* :3}
    *Maximal Frequent Candidate Table* contains the items, {*abcd* : 2}
After mining segment 1,
    *Maximal Frequent Table* contains the itemsets {*abc* : 4}
    *Maximal Frequent Candidate Table* contains the items,{*ade* : 1, *abce* : 1, *abcd* : 2}
After mining segment 3,
    *Maximal Frequent Table* contains the itemsets {*abcd* : 3}

*Maximal Frequent Candidate Table* contains the items, {*ade* : 1, *abce* : 2}

Step II:

Common frequent patterns in *Maximal Frequent Candidate Table* {*ae* : 3}
    *Maximal Frequent Table* contains the itemsets {*abcd* : 3, *ae* : 3}
Finally all the MFS are available in *Maximal Frequent Table*.
    Thus it is seen that the proposed technique generates the same MFS as those generated by the Pincer approach of (Lin and Kedem, 2002).

## 5. PERFORMANCE EVALUATION

This section gives details of the experimental evaluation of the performance of the proposed algorithm.

### 5.1 EXPERIMENTAL SETUP AND RESULTS

The experiment was conducted on a Pentium IV computer with a CPU clock rate of 2.8 GHz, 2 GB RAM running Windows Operating System. The data sources for the experiment were generated synthetically. Three data sources 10K, 50K, and 100K were generated to study the performance of the proposed work. Table 5 shows the characteristics of each data source. The performance of the proposed approach has extensively been studied to confirm its effectiveness.

**Table 5 Summary of sample data sources taken for performance study**

| Name | |T| | |I| | |D| |
|---|---|---|---|
| Data source 1 | 10 | 3 | 10K |
| Data source 2 | 10 | 3 | 50K |
| Data source 3 | 10 | 3 | 100K |

The MFS generated by the proposed technique has been found to be identical to the MFS generated by the Pincer Approach.

Tables 6, 7, and 8 simply compare the performance of the proposed technique with the Pincer technique. Figure 3 displays the diagrammatic representation of these results.

**Table 6 Performance comparison on data source T10.I3.10K**

| Technique | Supports | | | | |
|---|---|---|---|---|---|
| | 10 | 20 | 30 | 40 | 50 |
| Pincer (Time in ms.) | 126 | 105 | 104 | 93 | 80 |
| Proposed (Time in ms.) | 87 | 85 | 79 | 74 | 61 |

**Table 7 Performance comparison on data source T10.I3.50K**

| Technique | Supports | | | | |
|---|---|---|---|---|---|
| | 10 | 20 | 30 | 40 | 50 |
| Pincer (Time in sec.) | 115 | 110 | 79 | 75 | 73 |
| Proposed (Time in sec.) | 101 | 99 | 73 | 67 | 56 |

**Table 8 Performance comparison on data source T10.I3.100K**

| Technique | Supports | | | | |
|---|---|---|---|---|---|
| | 10 | 20 | 30 | 40 | 50 |
| Pincer (Time in sec.) | 327 | 312 | 290 | 245 | 238 |
| Proposed (Time in sec.) | 276 | 270 | 268 | 221 | 193 |

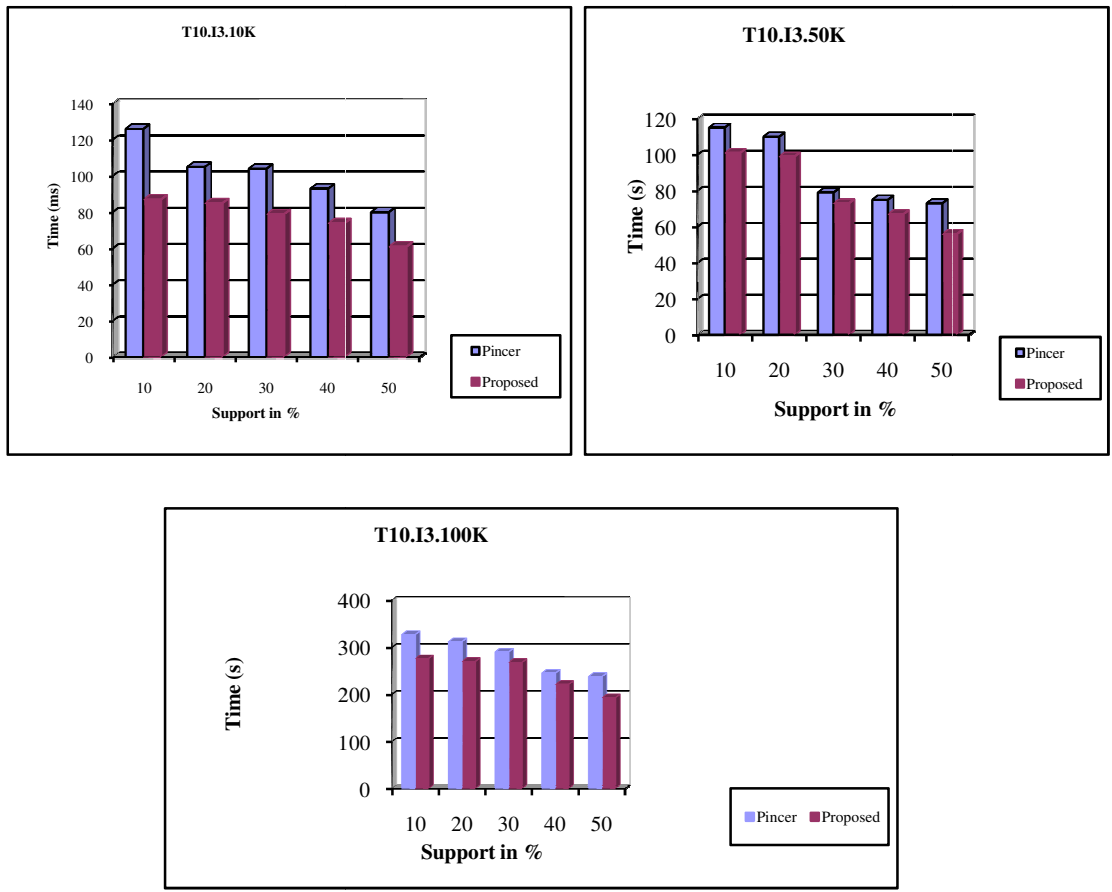

**Figure 3  Performance evaluation**

From the simulation results, the proposed method takes minimum time compared to the Pincer technique. All these experiments are conducted for the supports varying from 50% to 10%.  The time requirement to generate MFS on data source T10.I3.10K by the Pincer was from 80 ms to 126 ms and by the proposed method was from 61 ms to 87 ms. In the data source T10.I3.50K, the time requirement to generate MFS by the Pincer was from 73 seconds to 115 seconds and by the proposed method was from 56 seconds to 101 seconds.  For the third data source T10.I3.100K, the time requirement to generate MFS by the Pincer was from 238 seconds to 327 seconds and by the proposed method was from 193 seconds to 276 seconds. This shows that the proposed method performs better on dense data sources. From Tables 6, 7, and 8 and Figure 3, it can be seen that the time for generating MFS by the proposed technique is reduced by 10 to 20% when compared to the Pincer method.

## 6. CONCLUSION

Maximal Frequent Set (MFS) generation is identified as a time consuming problem. Different techniques and methodologies exist to generate MFS. The proposed technique mines MFS with the help of simple data structures. It uses partitioning approach to guide the search. It requires only two passes over the transactional data source. All this results in reduction in time for the MFS generation by about 10-20% in comparison with the Pincer method.